\documentclass{jfm}

\usepackage{graphicx}
\usepackage{epstopdf, epsfig}
\usepackage{amsmath}
\usepackage{amssymb}
\usepackage{commath}
\usepackage{gensymb}
\usepackage[percent]{overpic}
\usepackage{color}
\usepackage{bm}
\usepackage{amsfonts}
\usepackage{comment}
\usepackage{soul}
\usepackage{mleftright}

\newcommand{\D}{\mathrm{d}}
\newcommand{\sgn}{\operatorname{sgn}}

\definecolor{darkspringgreen}{rgb}{0.09, 0.65, 0.27}

\shorttitle{Effect of a surface tension imbalance on a partly submerged cylinder}
\shortauthor{S.\ D.\ Janssens, V.\ Chaurasia and E.\ Fried}

\title{Effect of a surface tension imbalance on a partly submerged cylinder}

\author{Stoffel D. Janssens\aff{1},
  Vikash Chaurasia\aff{1}$^,$\aff{2}
 \and Eliot Fried\aff{1}\corresp{\email{eliot.fried@oist.jp}}}

\affiliation{\aff{1}Okinawa Institute of Science and Technology Graduate University (OIST), Onna, Okinawa 904-0495, Japan
\aff{2}Department of Mechanical Engineering, University of Houston, Houston, TX 77004, USA}

\begin{document}

\maketitle

\begin{abstract}
We perform a static analysis of a circular cylinder that forms a barrier between surfactant-laden and surfactant-free portions of a liquid--gas interface. In addition to determining the general implications of the balances for forces and torques, we quantify how the imbalance $\Delta\gamma=\gamma_a-\gamma_b$ between the uniform surface tension $\gamma_a$ of the surfactant-free portion of the interface and the uniform surface tension $\gamma_b$ of the surfactant-laden portion of the interface influences the load-bearing capacity of a hydrophobic cylinder. Moreover, we demonstrate that the difference between surface tensions on either side of a cylinder with a cross-section of arbitrary shape induces a horizontal force component $f^h$ equal to $\Delta \gamma$ in magnitude, when measured per unit length of the cylinder. With an energetic argument, we show that this relation also applies to rod-like barriers with cross-sections of variable shape. In addition, we apply our analysis to amphiphilic Janus cylinders and we discuss practical implications of our findings for Marangoni propulsion and surface pressure measurements.
\end{abstract}

%\keywords surface pressure, floating, sinking, Langmuir--Blodgett trough

\section{Introduction}
\label{Section:Introduction}
\begin{figure}
\centering
\begin{overpic}{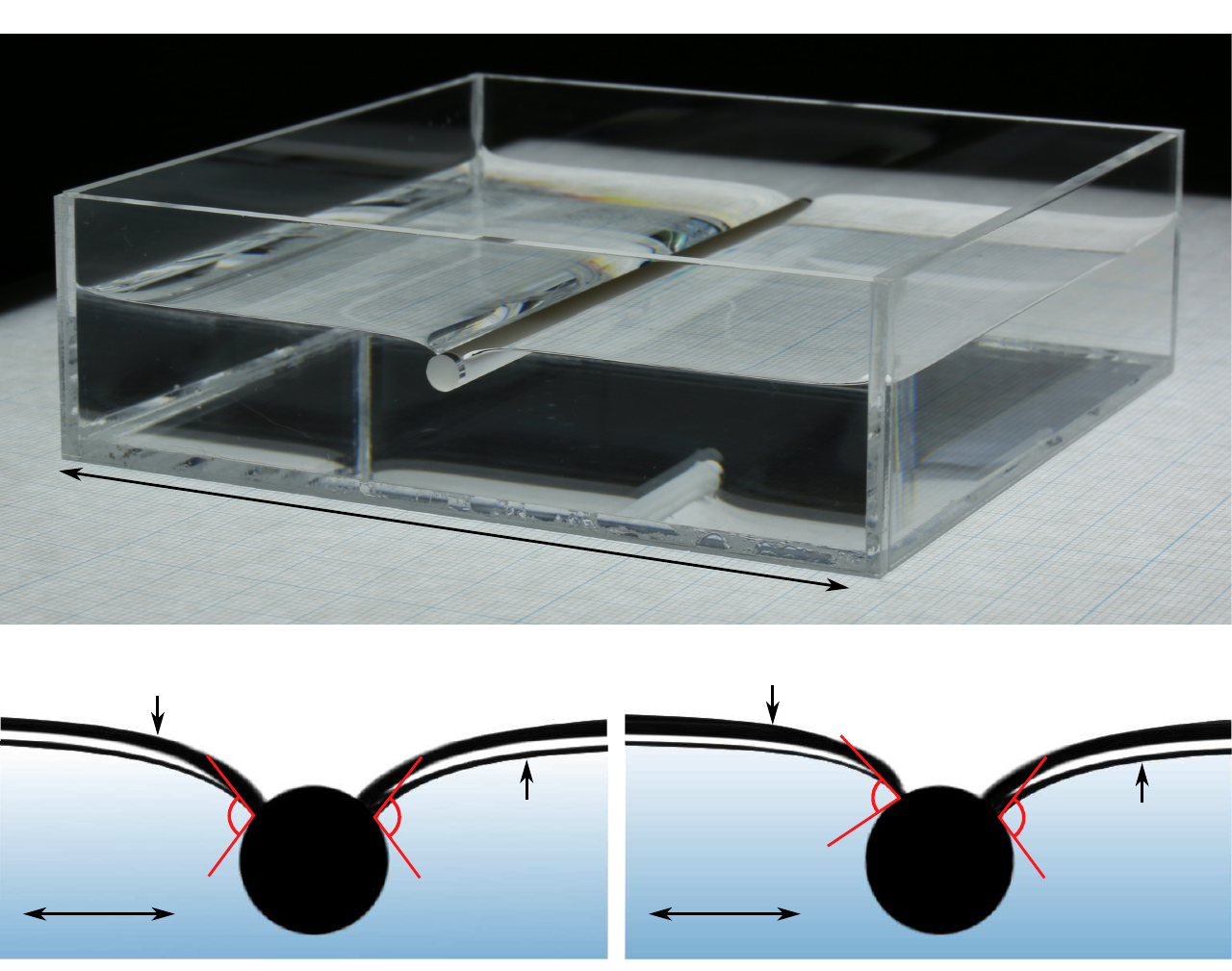}
\put(1,73){\color{white}($a$)}%
\put(30,34){\rotatebox{-8}{10 cm}}
%\put(71,38){\rotatebox{65}{10 cm}}
%\put(65,38){\color{white}\rotatebox{89}{3 cm}}
\put(1,25){($b$)}
\put(5,5.5){4 mm}
\put(41,23){air}
\put(39,5){water}
\put(21,9){\color{white}cylinder}%
\put(8,23.5){wetting of bath wall}
\put(39,12){interface}
\put(33,11.75){$\theta_a$}
\put(16,11.75){$\theta_b$}
\put(51,25){($c$)}
\put(55.5,5.5){4 mm}
\put(91,23){air}
\put(89,5){water}
\put(72,9){\color{white}cylinder}%
\put(58,24){wetting of bath wall}
\put(88.5,12){interface}
\put(83.75,11.75){$\theta_a$}
\put(68.25,13.75){$\theta_b$}
\end{overpic}
\caption{($a$) Photograph of an acrylic glass water bath ($10 \times 10 \times 3$~cm$^3$), resting on a light box, to which a partly submerged polytetrafluoroethylene (PTFE) cylinder with diameter 4.2~mm and length 10~cm is attached. ($b$) Side view photograph, taken with a drop shape analyzer (DSA 100 Kr\"uss), of the cylinder in ($a$). The gradient and the angles are superimposed on the original photograph for clarity. The thin, lowermost curves are formed by the interaction of the water and the cylinder, and the thick, uppermost curves result from the wetting of the bath wall. On both sides of the cylinder, the surface tensions are approximately 72~mN/m, the value for a water--air interface. The contact angles $\theta_a$ and $\theta_b$ are essentially equal and greater than $\upi/2$ ($\theta_a \cong \theta_b > \upi/2$), which is consistent with the hydrophobic character of PTFE. ($c$) Side view photograph, again taken with a drop shape analyzer (DSA 100 Kr\"uss), of the cylinder as in ($a$); however, the surface tension $\gamma_b$ on the left-hand side of the cylinder is lowered relative to that, $\gamma_a$, on the right-hand side of the cylinder by drop casting a practically water insoluble surfactant (oleic acid). The contact angles on either side of the cylinder obey $\theta_b<\upi/2<\theta_a$ and the profile of the interface is no longer symmetric.}
\label{experiment}
\end{figure}
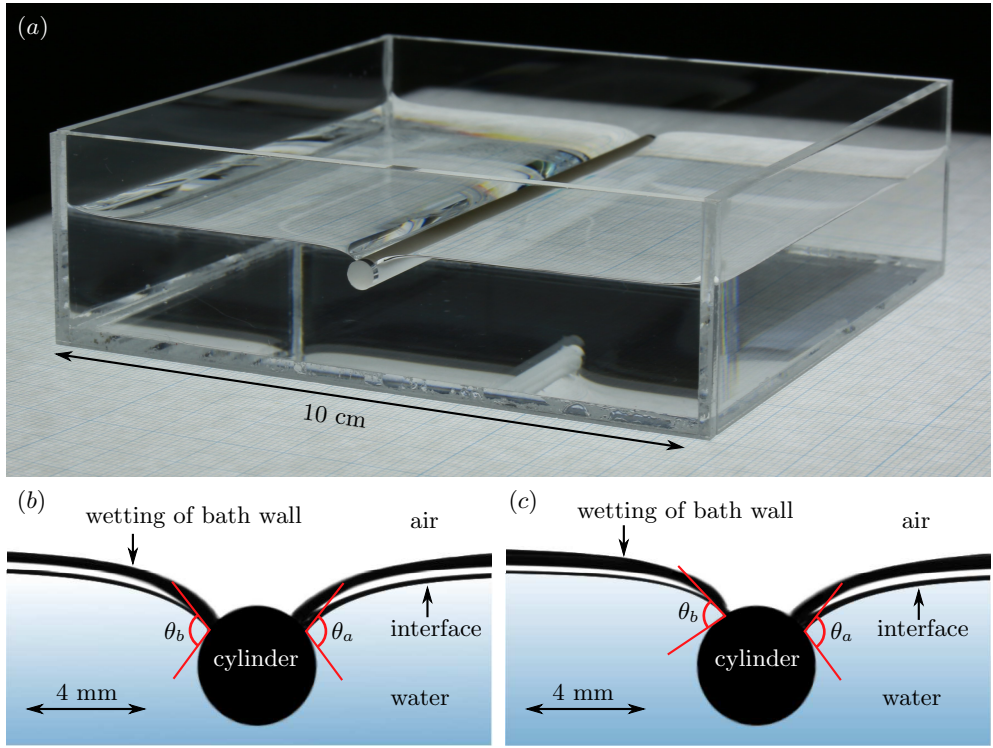
According to \cite{Lauga2012}, Marangoni propulsion \citep{Bush2006,Masoud2014} refers to `the situation where a body, located at the surface of a fluid, generates an asymmetric distribution of surface active materials, thereby prompting a surface tension imbalance and a Marangoni flow, both of which lead to locomotion.' Examples of such bodies, which are sometimes called Marangoni surfers \citep{Wurger2014,Vandadi2017}, include camphor scrapings \citep{VanderMensbrugghe1869,Nakata1997}, organic solvent droplets \citep{Oshima2014,Janssens2017}, and water walking insects \citep{Linsenmair1963}.

In this work, we explore how interfacial curvature influences Marangoni propulsion, a concern recently raised by \cite{Vandadi2017} and \cite{Janssens2017}. To focus on curvature, we consider a situation where a surface tension imbalance is present but a Marangoni flow is absent. Experimentally, this situation is realized by the Langmuir film balance \citep{Langmuir1917}, in which a barrier separates surfactant-laden and surfactant-free portions of a liquid--gas interface. These respective interfaces have uniform surface tensions $\gamma_a$ and $\gamma_b$ that generate a surface tension imbalance $\Delta \gamma = \gamma_a - \gamma_b$.  Therefore, we consider a class of problems inspired by the set-up present in such a balance. The horizontal component $f^h$ of the force that acts on a barrier, due to $\Delta \gamma$, is measured with the balance. If the liquid--gas interface is assumed to be flat, then $f^h$ is equal in magnitude to $\Delta \gamma$ when measured per unit length of the barrier. In actuality, however, the interface is curved (figure~\ref{experiment}) and this appears, at first glance, to complicate the calculation of $f^h$. In \S\ref{Section:Static_analysis}, we therefore perform a static force and torque analysis of a circular cylinder that acts as a surfactant barrier. This analysis, which generalizes studies of cylinders floating on liquids with uniform surface tension performed by \citet{Princen1969}, \citet{Rapacchietta1977}, \citet{Bhatnagar2006}, \citet{Vella2006}, and \citet{Liu2007}, has never (at least to our knowledge) previously appeared in the literature.

One important consequence of our analysis is that lowering the surface tension on one side of a floating cylinder can alter the load-bearing capacity of that cylinder. In \S\ref{Subsection:Demonstration_of_vertical_force_analysis}, we therefore conduct a vertical force analysis on a hydrophobic cylinder. For more information on how objects at a liquid--gas interface with uniform surface tension float, we refer the reader to work by \cite{McCuan2013} and to a review by \citet{Vella2015}. In \S\ref{Subsection:Demonstration_of_horizontal_force_analysis}, we perform a horizontal force analysis from which we infer that the magnitude of $f^h$ is equal to $\Delta \gamma$. In \S\ref{Subsection:Janus_cylinder}, we extend the force analysis to the problem of a floating amphiphilic Janus cylinder and compare our results with the work on Janus beads by \citet{Casagrande1989}.

In \S\ref{Section:Energetic_argument} we use an energetic argument to prove that the magnitude of $f^h$ is equal to $\Delta \gamma$ and show that this relation is applicable to rod-like barriers with cross-sections of variable shape.

Lastly, in \S\ref{Section:Practical_implications}, we place our findings into context and discuss practical implications of the analysis for Marangoni propulsion and surface pressure measurements.

\section{Static analysis}
\label{Section:Static_analysis}

\subsection{Setup of the problem}
\label{Section:Setup}

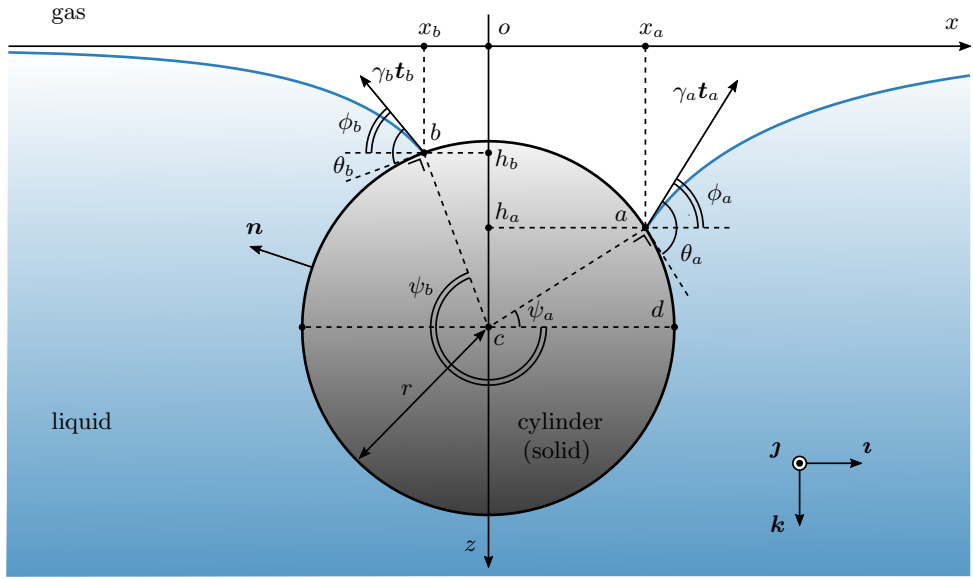
\begin{figure}
\centering
\begin{overpic}{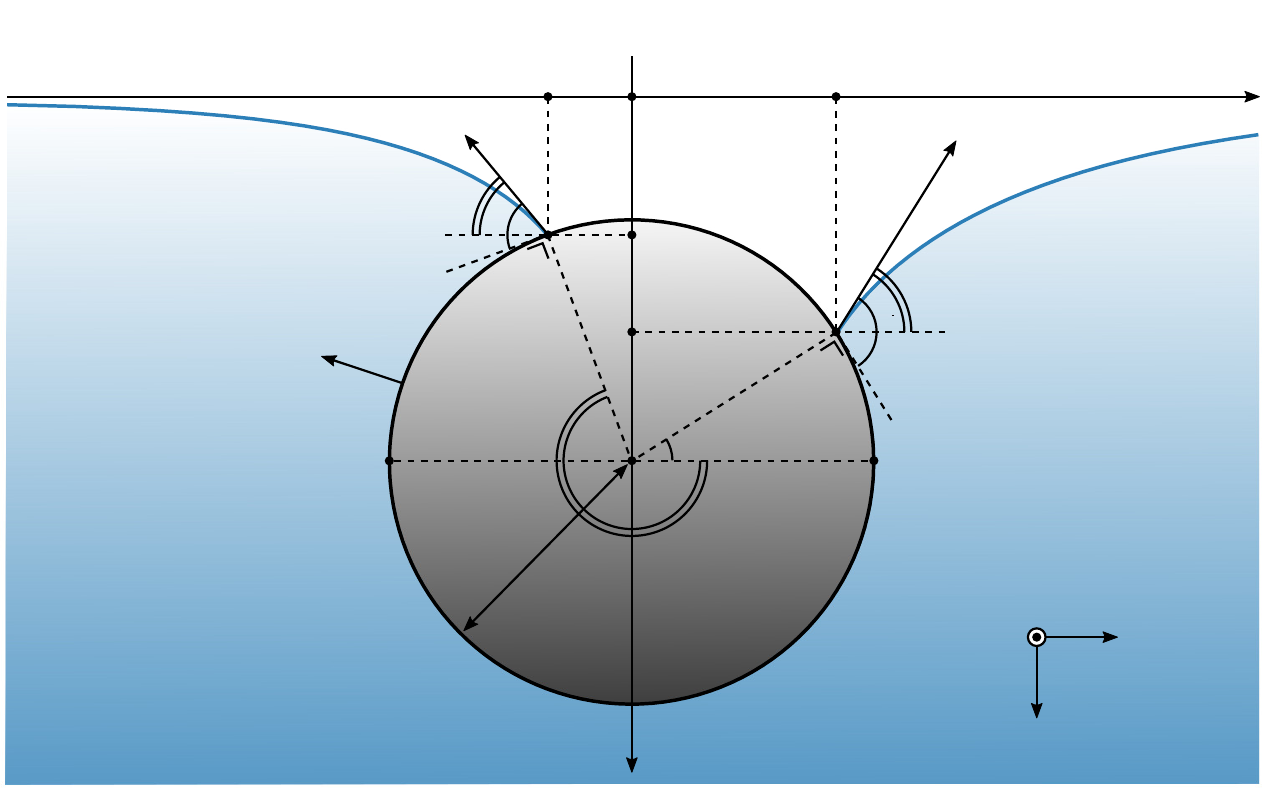}
   \put(97,57){$x$}%
   \put(47.5,3.5){$z$}%
   \put(88.5,13.5){$\bm{\imath}$}%
   \put(79,13.5){$\bm{\jmath}$}%
   \put(79,5.5){$\bm{k}$}%
   \put(51,56.5){$o$}%
   \put(41,19.5){$r$}%
   \put(63,37.35){$a$}%
   \put(65.5,56.5){$x_a$}%
   \put(44,45.6){$b$}%
   \put(42.75,56.5){$x_b$}%
   \put(50.5,24.6){$c$}%
   \put(66.75,27.5){$d$}%
   \put(25,36){$\bm{n}$}%
   \put(54,27.5){$\psi_{a}$}%
   \put(41.75,30){$\psi_{b}$}%
   \put(50.65,37.5){$h_a$}%
   \put(50.65,43){$h_b$}%
   %\put(28.25,32){\color[rgb]{0,0,0}\makebox(0,0)[lb]{\smash{$\abs{h_c}$}}}%
   \put(69,50){$\gamma_a \bm{t}_a$}%
   \put(38,52){$\gamma_b \bm{t}_b$}%
   \put(72.5,39.5){$\phi_a$}%
   \put(69.75,33.25){$\theta_a$}%
   \put(34.5,46.75){$\phi_b$}%
   \put(34,42.25){$\theta_b$}%
   \put(53,16){cylinder}%
   \put(53.5,13){(solid)}%
   \put(5,16){\color[rgb]{0,0,0}\makebox(0,0)[lb]{\smash{liquid}}}%
   \put(5,58){\color[rgb]{0,0,0}\makebox(0,0)[lb]{\smash{gas}}}%
   %\put(2,50.5){$c = (0,h_c)$}%
   %\put(2,60.5){$a = (x_a,h_a)$}%
   %\put(2,55.5){$b = (x_b,h_b)$}%
\end{overpic}
\caption{Schematic of a circular cylinder of radius $r$ laying at a liquid--gas interface. The cylinder acts as a barrier between surfactant-free and surfactant-laden portions of that interface. The respective interfaces have uniform surface tensions $\gamma_a$ and $\gamma_b$. Far from the cylinder, both interfaces are flat and have the same horizontal elevation. The orthonormal basis vectors $\bm{\imath}$, $\bm{\jmath}$, and $\bm{k}$ correspond to a rectangular Cartesian coordinate system with origin $o$. The centre of the cylinder is horizontally aligned with $o$ and is located at point $c$ with coordinates $x = 0$ and $z=h_c$. The undisturbed portions of the liquid--gas interface are at the same height as $o$. The angles $\psi_a$ and $\psi_b$ define a circular arc and are measured clockwise, with reference to the figure, starting from ray $cd$ and satisfy $\psi_a<\psi_b$. Therefore, $\psi_a$ is negative for the situation depicted in the figure. The end points of that arc are located at solid--liquid--gas contact lines $a$ and $b$. Point $a$ has coordinates $x = x_a$ and $z = h_a$, and point $b$ has coordinates $x = x_b$ and $z = h_b$. The contact angles $\theta_a$ and $\theta_b$ are measured in the liquid phase and are always positive. The unit tangent vectors $\bm{t}_a$ and $\bm{t}_b$, of the liquid--gas interface at $a$ and $b$, respectively, are at respective angles $\phi_a$ and $\phi_b$ relative to the horizon. For the situation depicted in the figure, $\phi_a$ and $\phi_b$ are positive. The unit normal vector $\bm{n}$ points from the surface of the cylinder into the liquid and gas.}
\label{ForceDiagramCylinder}
\end{figure}

Figure~\ref{ForceDiagramCylinder} shows a schematic of a circular cylinder of radius $r$ acting as a surfactant barrier and illustrates the problem we address in this work. The surfactant-free and surfactant-laden portions of the interface have uniform surface tensions $\gamma_a$ and $\gamma_b$. Edge effects can be ignored if the cylinder is much longer than the capillary length
\begin{equation}
   l_c = \sqrt{\frac{\gamma_a}{\rho_lg}},
   \label{CapillaryLength}
   \end{equation}
where $\rho_l$ and $g$ denote the mass density of the liquid and the gravitational acceleration on earth. For our purposes, it suffices to take the cylinder to be of infinite length.

The orthonormal basis vectors $\bm{\imath}$, $\bm{\jmath}$, and $\bm{k}$ are chosen so that $\bm{k}$ is parallel to the direction along which gravity acts, $\bm{\imath}$ is directed from the surfactant-laden portion of the liquid--gas interface to the surfactant-free portion of that interface, and $\bm{\jmath}$ is parallel to the cylinder axis. The Cartesian coordinates $x$ and $z$ increase in the directions of $\bm{\imath}$ and $\bm{k}$, respectively. Far from the cylinder, both portions of the liquid--gas interface have the same horizontal elevation and the origin $o$ is positioned at that elevation. The axis of the cylinder is located at point $c$ with coordinates $x = 0$ and $z = h_c$. The lines along which the surfactant-free and surfactant-laden portions of the liquid--gas interface meet the cylinder are denoted by points $a$ and $b$, respectively, which are at respective angles $\psi_a$ and $\psi_b$. These angles are measured clockwise, with reference to figure~\ref{ForceDiagramCylinder}, starting from ray $cd$ and satisfy $\psi_a < \psi_b$. Thus, if $a$ is located above $cd$, which corresponds to the situation depicted in figure~\ref{ForceDiagramCylinder}, $\psi_a < 0$. However, if $a$ is located below $cd$, $\psi_a > 0$. The wetted portion of the cylinder surface is found by following the surface of the cylinder clockwise from $a$ to $b$. We express the respective $x$-coordinates $x_a$ and $x_b$ of $a$ and $b$ as
\begin{equation}
x_a = r\cos{\psi_a} \qquad \text{and} \qquad x_b = r\cos{\psi_b},
\label{x}
\end{equation}
and the respective $z$-coordinates $h_a$ and $h_b$ of $a$ and $b$ as
\begin{equation}
h_a = r \sin{\psi_a} + h_c \qquad \text{and} \qquad h_b = r \sin{\psi_b} + h_c.
\label{h}
\end{equation}
The contact angles $\theta_a$ and $\theta_b$ are measured in the liquid and are always positive. The unit tangent vectors $\bm{t}_a$ and $\bm{t}_b$ of the liquid--gas interface at $a$ and $b$ are at respective angles $\phi_a$ and $\phi_b$ relative to the horizon. These angles are positive for the situation depicted in figure~\ref{ForceDiagramCylinder}. For reference, the various angles are connected by the geometrical relations 
\begin{equation} 
	\phi_a = \theta_a - \psi_a - \frac{\upi}{2} \qquad \text{and} \qquad
	\phi_b = \theta_b + \psi_b - \frac{3\upi}{2}.
	\label{georel}
	\end{equation}
The surfactant-laden and surfactant-free portions of the liquid--gas interface are represented by
\begin{equation}
\{(x,z):-\infty<x\le x_b,z=h(x)\} \qquad \text{and} \qquad \{(x,z):x_a\le r<+\infty,z=h(x)\},
\end{equation}
respectively.
Finally, the unit normal to the cylinder surface, directed outward into liquid and gas, is denoted by $\bm{n}$. 

In contrast to the setup in figure~\ref{experiment}, we allow for vertical translation and the possibility of a uniform line load $\bm{f}_L$ on the apex of the cylinder. Horizontal translation of the cylinder is prevented by a reactive horizontal force $\bm{f}_R$. We represent $\bm{f}_R$ by a traction that includes components that act horizontally and normal to the surface of the cylinder. The net force $\bm{f}_N$ acting on the cylinder is zero in magnitude and, in addition to $\bm{f}_L$ and $\bm{f}_R$, accounts for the weight $\bm{f}_G$ of the cylinder, the force $\bm{f}_P$ due to the hydrostatic pressure, and the force $\bm{f}_T$ due to surface tension. Rotation of the cylinder about its axis is prevented by a reactive torque $\bm{\tau}_R$ induced by a pair of horizontal forces that are equal in magnitude but opposite in direction and act at the apex and the base of the cylinder surface. All forces, torques, and areas considered in this work are measured per unit length of the cylinder.

\subsection{Weight and hydrostatic pressure}

\label{Subsection:Weight_and_hydrostatic_pressure_force}

\subsubsection{Weight}

Consistent with the convention that $z$ increases in the direction $\bm{k}$ along which gravity acts, we write
\begin{equation}
   f_G=|\bm{f}_G|= \rho_c  g \upi  r^2,
   \label{Gravity}
   \end{equation}
where $\rho_c$ and $r$ denote the mass density and the radius of the cylinder, respectively. It is convenient to let $\rho_c^*$ denote the maximum allowed mass density for which the cylinder floats at the liquid--gas interface in the absence of a line load $\bm{f}_L$.

\subsubsection{Hydrostatic pressure}

The force $\bm{f}_{P}$ due to hydrostatic pressure $p$ is obtained by integrating $-p\bm{n}$ over the part $\mathcal{A}$ of the cylinder surface that is in contact with the liquid. This operation gives
 \begin{equation}
    \bm{f}_P  = - \int_\mathcal{A}p \bm{n}\,\D s,
    \label{PressureForce}
    \end{equation}
where $\bm{n}$ is the unit normal to the surface of the cylinder, directed into the liquid, and $\D s$ is the surface element. We assume that the pressure in the gas is uniform and take $p$ equal to zero at the horizontal level of the liquid--gas interface in the far field on either side of the cylinder.

The generalized Archimedes principle \citep{Mansfield1997,Keller1998} can be applied to obtain a more specific relation for the vertical component $f_P^v = \bm{f}_P \bm{\cdot} \bm{k}$ of $\bm{f}_P$. The surface tension difference $\Delta \gamma$ induces an asymmetry which is confirmed by the experiment shown in figure~\ref{experiment}(c), so that the $z$-coordinates $h_a$ and $h_b$ of the respective solid--liquid--gas contact lines at $a$ and $b$ are not equal. The horizontal component $f_P^h = \bm{f}_P \bm{\cdot} \bm{\imath}$ of $\bm{f}_P$ is therefore generated by an average pressure $\bar{p}$ that acts over a projected (signed) area $\Delta h = h_a - h_b$, giving
\begin{equation}
   f_P^h = \bar{p} \Delta h = \bar{p} (h_a - h_b).
   \label{HorizontalGeneral}
   \end{equation}
Introducing $\bar{p} = \rho_l g h_m$, with $h_m = (h_a + h_b)/2$, we rewrite relation \eqref{HorizontalGeneral} as
\begin{equation}
   f_P^h = \frac{\rho_l g}{2}(h_a^2 - h_b^2).
   \label{HorizontalGeneralFinal}
   \end{equation} 

Following \cite{Keller1998}, we obtain \eqref{HorizontalGeneralFinal} by parametrizing $\mathcal{A}$ with  respect to arclength $t$, from $b$ to $a$, so that $\D s = \D t$, $t_b = 0$, and $t_a = \mathcal{A}$. Using a superposed hat to indicate parametrization with respect to $t$ and denoting the derivative with respect to $t$ with a subscript, we then express $p$ and $\bm{n}$ respectively as
\begin{equation}
   p = \rho_lg\hat{z} \qquad \text{and} \qquad \bm{n} = -\hat{z}_t \bm{\imath} + \hat{x}_t \bm{k}.
   \label{pandn}
   \end{equation}
Writing $f_{P}^h = \bm{f}_P \bm{\cdot} \bm{\imath}$, and using \eqref{PressureForce} and \eqref{pandn}, we thus obtain
\begin{align}
   f_P^h &= - \rho_lg\int_{t_b}^{t_a} \hat{z} \mleft(-\hat{z}_t \bm{\imath} + \hat{x}_t \bm{k}\mright)\bm{\cdot \imath}\, \D t\notag\\
   &= \rho_lg\int_{h_b}^{h_a} z\, \D z \notag\\
   &= \frac{\rho_lg}{2}(h_a^2 - h_b^2).
   \end{align}
It is noteworthy that relation \eqref{HorizontalGeneralFinal} is valid for cylinders of noncircular cross-section.

For a cylinder with circular cross-section, we can also write $\D s = r\,\D{\psi}$, and using \eqref{PressureForce}, obtain
 \begin{equation}
    \bm{f}_P  =- r \int_{\psi_a}^{\psi_b} p \bm{n}\,\D \psi,
    \label{PressureForceCylinder}
    \end{equation}
where $\psi_a$ and $\psi_b$ are the angles defining a circular arc with end points located at the contact lines $a$ and $b$. For a point on the cylinder surface at elevation $r\sin{\psi}+h_c$, where $h_c$ is the $z$-coordinate of the centre $c$ of the cylinder, we may express $p$ as
\begin{equation}
   p = \rho_l g \mleft(r \sin{\psi} + h_c \mright).
   \label{Pressure}
   \end{equation}
Writing $\bm{n} \bm{\cdot} \bm{k} = \sin{\psi}$ and $\bm{n} \bm{\cdot} \bm{\imath} = \cos{\psi}$, we thus find that the respective vertical and horizontal components $f_{P}^v$ and $f_{P}^h $ of $\bm{f}_P$ are given by
 \begin{align}
  f_P^v &= -\rho_l g r \int_{\psi_a}^{\psi_b} \mleft(r \sin{\psi} + h_c\mright) \sin{\psi}\,\D{\psi}
\notag\\[4pt]  
   &= - \frac{\rho_l g  r^2}{2}\mleft(\Delta\psi + \sin{\psi_a}  \cos{\psi_a} - \sin{\psi_b}  \cos{\psi_b}
    + \frac{2h_c}{r} \mleft(\cos{\psi_a}-\cos{\psi_b}\mright)\mright),
    \label{PressureForceCylinderZI}
    \end{align}
with $\Delta \psi = \psi_b - \psi_a$, and
   \begin{align}
    f_P^h &= - \rho_l g r \int_{\psi_a}^{\psi_b} \mleft(r \sin{\psi} + h_c\mright) \cos{\psi}\,\D{\psi}
\notag\\[4pt]
    &
    = \frac{\rho_l g r^2 \mleft(\sin{\psi_a}-\sin{\psi_b}\mright)}{2}
    \mleft(\sin{\psi_a} + \sin{\psi_b} + \frac{2h_c}{r} \mright).
    \label{PressureForceCylinderXI}
    \end{align}
The relations \eqref{PressureForceCylinderZI} and \eqref{PressureForceCylinderXI} can also be obtained as respective consequences of \eqref{HorizontalGeneralFinal} and the generalized Archimedes principle \citep{Mansfield1997,Keller1998}.

\subsection{Surface tension}
\label{Subsection:Surface_tension_force}

The unit tangent vectors $\bm{t}_a$ and $\bm{t}_b$, of the liquid--gas interface at the contact lines $a$ and $b$, are at respective angles $\phi_a$ and $\phi_b$ relative to the horizon. The force $\bm{f}_T$ due to the surface tension is the sum of $\gamma_a \bm{t}_a $ and $\gamma_b \bm{t}_b$. The vertical and horizontal components $f_T^v=\bm{f}_T\bm{\cdot}\bm{k}$ and $f_T^h=\bm{f}_T\bm{\cdot}\bm{\imath}$ of $\bm{f}_T$ are given by
\begin{equation}
    f_T^v = - \gamma_a  \sin{\phi_a} -  \gamma_b  \sin{\phi_b}
\qquad\text{and}\qquad 
    f_T^h = \gamma_a  \cos{\phi_a} - \gamma_b  \cos{\phi_b}.
   \label{TensionForces}
    \end{equation}

\subsection{Torque}

The torque $\bm{\tau}$ about the axis of the cylinder is the sum of $\bm{\tau}_a = r\bm{n} \times \gamma_a\bm{t}_a$ and $\bm{\tau}_b = r\bm{n} \times \gamma_b\bm{t}_a$. The only nonvanishing component of $\bm{\tau}$, namely $\tau=\bm{\tau}\bm{\cdot}\bm{\jmath}$, is given by
\begin{equation}
    \tau = r \mleft(\gamma_b  \cos{\theta_b} - \gamma_a  \cos{\theta_a} \mright),
    \label{Torque}
\end{equation}
where $\theta_a$ and $\theta_b$ are the respective contact angles at the contact lines $a$ and $b$, measured in the liquid phase. Because $\bm{f}_P$ and $\bm{f}_G$ act radially, they do not exert any torque on the cylinder about its axis. As a result, the net torque $\bm{\tau}_N$ is $\tau=\bm{\tau}\bm{\cdot}\bm{\jmath}$. This quantity is balanced by $\bm{\tau}_R$, which prevents rotation of the cylinder about its axis.  Writing $\tau_R = \bm{\tau}_R \bm{\cdot \jmath}$, we express this balance as $-\tau = \tau_R$. \cite{Singh2004} showed that the torque vanishes if the contact angles and the surface tensions are the same. By specializing \eqref{Torque} appropriately, we recover this result.

\subsection{Net force on cylinder}
\label{Subsection:Components_of_the_net_force}

Let $\bm{f}$ be the sum of $\bm{f}_P$, $\bm{f}_T$, and $\bm{f}_G$, with respective vertical and horizontal components 
\begin{equation}
\label{FVFN}
f^v = \bm{f \cdot k} = f_P^v + f_T^v + f_G \qquad \text{and} \qquad f^h = \bm{f \cdot \imath} = f_P^h + f_T^h ,
\end{equation}
where, with reference to \eqref{Gravity}, \eqref{PressureForceCylinderZI}, and \eqref{TensionForces}$_1$, $ f^v$ is given by
\begin{multline}
     f^v = - \frac{\rho_l g  r^2}{2}\mleft(\Delta\psi + \sin{\psi_a}  \cos{\psi_a} - \sin{\psi_b}  \cos{\psi_b} +\frac{2h_c}{r} \mleft(\cos{\psi_a} - \cos{\psi_b}\mright)\mright)
     \\[4pt]
     - \gamma_a  \sin{\phi_a} -  \gamma_b  \sin{\phi_b} + \rho_c g\upi  r^2,
\label{fvFull}
\end{multline}
and, with reference to \eqref{PressureForceCylinderXI} and \eqref{TensionForces}$_2$, $ f^h$ is given by
\begin{equation}
     f^h = \frac{\rho_l g r^2 \mleft(\sin{\psi_a}-\sin{\psi_b}\mright)}{2} \mleft(\sin{\psi_a} + \sin{\psi_b} + \frac{2h_c}{r} \mright) + \gamma_a  \cos{\phi_a} - \gamma_b  \cos{\phi_b}.
   \label{fhFull}
\end{equation}
Then, writing $f_L = \bm{f}_L \bm{\cdot k}$ and $f_R = \bm{f_}R \bm{\cdot \imath}$, we see that the vertical and horizontal components $f_N^v = \bm{f}_N \bm{\cdot k}$ and $f_N^h = \bm{f}_N \bm{\cdot \imath}$ of the net force $\bm{f}_N$ take the form
\begin{equation}
\label{FN}
f_N^v = f^v + f_L  \qquad \text{and} \qquad f_N^h = f^h + f_R.
\end{equation}
In equilibrium, the condition of force balance yields $f^v = -f_L$ and $f^h = -f_R$.

\subsection{Contact angle versus surface tension}
\label{Subsection:Contact_angle_versus_surface_tension}

The surface of a solid is hydrophobic if the static contact angle $\theta$, formed at a surfactant-free solid--water--air contact line located at the surface of that solid, is greater than $\upi/2$ when measured in the water phase. Moreover, a solid is hydrophilic if $\theta$ is less than $\upi/2$. \citet{Bargeman1973} reported on the contact angles that water droplets loaded with soluble surfactants make with hydrophobic substrates. Their measurements show that $\gamma_b \cos{\theta_b}$ decreases with increasing $\gamma_b$ according to the relation
\begin{equation}
\gamma_b \cos{\theta_b} = - \alpha \gamma_b + \beta,
    \label{surf_cont1}
    \end{equation}
where $\alpha$ and $\beta$ are obtained by linear regression. It is noteworthy that $\alpha$ and $\beta$ remain constant as the surfactant concentration increases from zero up to the critical micelle concentration. From works of \citet{Bargeman1973}, \citet{Szymczyk2007} and \citet{Chaudhuri2012}, it is evident that $\alpha$ ranges from $0.6$ to $1.0$ and that $\beta$ ranges from 0 to 50~mN/m, depending on which surfactants and solids are involved. \citet{Lucassen-Reynders1963} showed that $\alpha$ can be represented by
\begin{equation}
\alpha = \frac{I_\textit{SL} - I_\textit{SG}}{I_\textit{LG}} = \frac{\D \mleft(\gamma_\textit{SL} - \gamma_\textit{SG}\mright)}{\D \gamma_\textit{LG}}
    \label{alpha}
    \end{equation}
where $I_{\textit{SG}}$, $I_{\textit{SL}}$, and $I_{\textit{LG}}$ denote the excess concentrations of surfactant molecules on the solid--gas, solid--liquid, the liquid--gas interfaces, respectively. This relation was established by invoking the equation of state \citep[p.~436]{Gibbs1878}
\begin{equation}
\frac{\D \gamma_i}{\D \ln \zeta} = - \mathcal{R}T_AI_i,
    \label{Gibbs}
    \end{equation}
where $i$ stands for $\textit{SL}$, $\textit{SG}$, or $\textit{LG}$, $\zeta$ is the activity of a surfactant, $\mathcal{R}$ is the ideal gas constant, and $T_A$ is the absolute temperature. Using the integrated form of \eqref{alpha} along with Young's~\citeyearpar{Young1805} equation,
\begin{equation}
\gamma_\textit{SG} - \gamma_\textit{SL} = \gamma_\textit{LG} \cos \theta,
    \label{Young}
    \end{equation}
results in \eqref{surf_cont1}. For $\alpha$ equal to unity, \citet{Bargeman1973} explained how, on the basis of Fowkes'~\citeyearpar{Fowkes1964} hypothesis, to find the constant of integration $\beta$ from independent experiments. With $\alpha$ and $\beta$ constant, and with $\gamma_b$ decreasing as a function of $I_{\textit{LG}}$, we see from \eqref{surf_cont1} that $\theta_b$ decreases with surfactant concentration.
 
Although our concern is with insoluble surfactants, we use \eqref{surf_cont1} as a first approximation. This is supported by the experiment shown in figure~\ref{experiment}(c), from which we find that $\theta_b$ decreases with surfactant concentration. From the same experiment, we also see that $\gamma_a$ and $\theta_a$ are not affected by changing $\gamma_b$ and, thus, $\theta_b$. A hydrophobic cylinder can therefore act as a surfactant barrier.

\subsection{Young--Laplace equation}
\label{Subsection:Young-Laplace_equations}

At the surfactant-free and surfactant-laden portions of the interface, force balance takes the form of the respective one-dimensional hydrostatic Young--Laplace equations \citep{Young1805,Laplace1805}
\begin{equation}
	\frac{\rho_l gh}{\gamma_a} = \frac{\sgn{\mleft(hh_{\textit{xx}}\mright)}h_{\textit{xx}}}{\mleft(1+ h_x^2\mright)^{3/2}} \qquad \text{and} \qquad
    	\frac{\rho_l gh}{\gamma_b} = \frac{\sgn{\mleft(hh_{\textit{xx}}\mright)}h_{\textit{xx}}}{\mleft(1+ h_x^2\mright)^{3/2}},
    \label{YL}
    \end{equation}
where a subscripted $x$ denotes differentiation with respect to $x$ and $\sgn$ denotes the signum function, so that the right-hand side of each equation represents the signed curvature of the appropriate portion of the liquid--gas interface. The relations in \eqref{YL} can be derived from the three-dimensional version $\Delta p = 2\gamma \mathcal{H}$ of the Young--Laplace equation, where $\gamma$ denotes the surface tension and $\mathcal{H}$ is the mean curvature of the surface.

\subsection{Scaling}
\label{Subsection:Nondimensionalization}

We adopt a scaling in which lengths are measured relative to $l_c$, mass densities are measured relative to $\rho_l$, forces are measured relative to $\gamma_a$, and torques are measured relative to $l_c\gamma_a$. In addition to the Bond number
\begin{equation}
\textit{Bo} = \mleft(\frac{r}{l_c}\mright)^2,
\label{Bond}
\end{equation}
this leads to dimensionless measures
\begin{equation}
\varGamma = \frac{\gamma_b}{\gamma_a}, 
\qquad \varGamma_0 = \frac{\beta}{\gamma_a},
\qquad D = \frac{\rho_c}{\rho_l},
\qquad\text{and}\qquad
D^* = \frac{\rho_c^*}{\rho_l},
\label{parametersND}
\end{equation}
of surface tension, the parameter $\beta$ entering the relation \eqref{surf_cont1} between $\gamma_b$ and $\theta_b$, mass density, and the maximum allowed mass density for which the cylinder floats. In addition, we introduce the dimensionless lengths
\begin{equation}
\{X, X_a, X_b, Z, R, H, H_a, H_b, H_c\} = \{x, x_a, x_b, z, r, h, h_a, h_b, h_c\}/l_c,
\label{ldim}
\end{equation}
%
%the 
dimensionless force components
\begin{equation}
\{F_N^v, F_N^h, F^v, F^h, F_P^h, F_T^h, F_L, F_R\} = \{f_N^v, f_N^h, f^v, f^h, f^h_P, f^h_T, f_L, f_R\}/\gamma_a,
\end{equation}
%
%and the 
%
and dimensionless torque components
\begin{equation}
\{T, T_R\} = \{\tau, \tau_R\}/(l_c \gamma_a).
\end{equation}
As a particular consequence of \eqref{Bond} and \eqref{ldim}, we see that $\textit{Bo} = R^2$. 

For the relations in \eqref{x} determining the $x$-coordinates $x_a$ and $x_b$ of $a$ and $b$, we obtain dimensionless counterparts
\begin{equation}
\label{X2}
X_a = R  \cos{\psi_a} \qquad \text{and} \qquad
X_b = R  \cos{\psi_b},
\end{equation}
and for the relations in \eqref{h} determining the $z$-coordinates $h_a$ and $h_b$ of $a$ and $b$, we obtain
\begin{equation}
    H_a = R  \sin{\psi_a} + H_c \qquad \text{and} \qquad
    H_b = R  \sin{\psi_b} + H_c.
    \label{Hc}
    \end{equation}
Furthermore, for the relations \eqref{HorizontalGeneralFinal}, \eqref{TensionForces}$_2$, \eqref{fvFull}, and \eqref{fhFull} determining the force components $f^h_P$, $f^h_T$, $f^v$, and $f^h$, we obtain
\begin{equation}
   F_P^h = \frac{1}{2}{(H_a^2 - H_b^2)},
    \label{HorizontalGeneralFinalND}
    \end{equation}
\begin{equation}
   F_T^h = \cos{\phi_a} - \varGamma \cos{\phi_b},
    \label{TensionForcesND}
    \end{equation}
\begin{multline}
    	F^v = - \frac{\textit{Bo}}{2} \mleft(\Delta\psi + \sin{\psi_a}  \cos{\psi_a} - \sin{\psi_b}  \cos{\psi_b} + \frac{2H_c}{R} \mleft(\cos{\psi_a}-\cos{\psi_b}\mright)\mright)
	\\[4pt]
- \mleft(\sin{\phi_a} + \varGamma \sin{\phi_b}\mright) + D \upi \textit{Bo},
\label{FvFull}
\end{multline}
and
\begin{equation}
     F^h = \frac{\textit{Bo}\mleft(\sin{\psi_a}-\sin{\psi_b}\mright)}{2}
     \mleft(\sin{\psi_a} + \sin{\psi_b} + \frac{2H_c}{R} \mright) + \cos{\phi_a} - \varGamma  \cos{\phi_b},
    \label{FhFull}
    \end{equation}
while for the relation \eqref{Torque} determining $\tau$, we obtain
\begin{equation}
   T = R (\varGamma \cos{\theta_b} - \cos{\theta_a}).
    \label{TorqueND}
    \end{equation}
Additionally, the relation \eqref{surf_cont1} connecting $\gamma_b$ and $\theta_b$ can be expressed in the form
\begin{equation}
   \theta_b = \arccos\mleft(-\alpha + {\frac{\varGamma_0}{\varGamma}}\mright),
    \label{surf_cont2}
    \end{equation}
while the equilibrium conditions \eqref{YL} yield
\begin{equation}
	H =\frac{\sgn{(HH_{X\mskip-3mu X})} H_{X\mskip-3mu X}}{(1+ H_{X}^{2})^{3/2}} \qquad \text{and} \qquad
    	\frac{H}{\varGamma} = \frac{\sgn{(HH_{X\mskip-3mu X})} H_{XX}}{(1+ H_{X}^{2})^{3/2}}.
    \label{YLND}
    \end{equation}

In \S\ref{Subsection:Demonstration_of_vertical_force_analysis}, we evaluate the effect of $\varGamma$ on the load-bearing capacity of a floating cylinder. We do this with the dimensionless quantity
\begin{equation}
\quad Q = \frac{\rho_c^* \mleft(\Delta \gamma\mright)}{\rho_c^* \mleft(0\mright)} = \frac{D^*\mleft(\varGamma\mright)}{D^*\mleft(1\mright)},
    \label{Q}
    \end{equation}
which is the ratio of $D^*$ in the presence of a surface tension imbalance to $D^*$ in the absence of a surface tension imbalance.

\subsection{Surface profiles}
\label{Subsection:Surface_Profiles}

\cite{Mansfield1997} observed that the results of integrating equations that are mathematically identical to those in \eqref{YLND}, with the same boundary conditions as in the present work, were expounded by \citet[p.~454--457]{Thompson1867}. We integrate \eqref{YLND}$_1$ and \eqref{YLND}$_2$  subject to the respective far-field conditions
\begin{equation}
\label{BCT}
\lim_{X\to +\infty} \{H, H_{X}\} \to 0 \qquad \text{and} \qquad \lim_{X\to -\infty} \{H, H_{X}\} \to 0.
\end{equation}
After the first integration we obtain
\begin{equation}
    \label{H}
   H_a = \sgn{\phi_a} \sqrt{2} \sqrt{1 - \cos{\phi_a}} \qquad \text{and} \qquad
   H_b = \sgn{\phi_b} \sqrt{2  \varGamma} \sqrt{1 - \cos{\phi_b}}.
    \end{equation}
If $\phi_a$ and $\phi_b$ are positive, as shown in figure~\ref{ForceDiagramCylinder}, then $H_a$ and $H_b$ should be positive. Alternatively, if $\phi_a$ and $\phi_b$ are negative, then $H_a$ and $H_b$ should be negative. The presence of signum functions in \eqref{H} assures this. After the second integration, we find that the surfactant-laden and surfactant-free profiles of the liquid--gas interface are given by
 \begin{equation}
  X = - \sqrt{4 - H^2} + \log \frac{2 + \sqrt{4 - H^2}}{\abs{H}} + C_a 
     \label{ProfileA}
\end{equation}
and 
\begin{equation}
  X = \sqrt{4 \varGamma - H^2} - \sqrt{\varGamma} \log \frac{2  \sqrt{\varGamma} + \sqrt{4  \varGamma - H^2}}{\abs{H}} - C_b,
     \label{ProfileB}
\end{equation}
respectively, where $C_a$ and $C_b$ denote constants that we determine from the respective near-field conditions \eqref{X2}.

\subsection{Specialization}
\label{Subsection:Specification}

If $\gamma_b=\gamma_a$, so that there is no surface tension imbalance and, thus, $\varGamma=1$, we write $\theta_a = \theta_b =\theta$, $\phi_a = \phi_b = \phi$, and $\psi_b = \upi - \psi_a$. By \eqref{FvFull} and \eqref{Hc}$_1$, we are thus led to a relation,
\begin{equation}
	F^v = -\textit{Bo}\mleft(\frac{\upi}{2} - \psi_a - \sin{\psi_a}\cos{\psi_a} + \frac{2H_a}{R} \cos{\psi_a}\mright) - 2 \sin{\phi} + D\upi\textit{Bo},
	\label{FvSym}
	\end{equation}
reminiscent of an earlier result due to \cite{Princen1969}. If, moreover, we completely neglect surface tension by taking $\gamma_b=\gamma_a=0$, then the liquid--gas interface must be flat. In this case, we write $H_a = 0$ and $\phi = 0$, so that, on setting $F^v = 0$ in \eqref{FvSym}, we arrive at a substantially simpler relation,
\begin{equation}
	\frac{1}{\upi}\mleft(\frac{\upi}{2} - \psi_a - \sin{\psi_a}\cos{\psi_a}\mright) = D,
	\label{FvArchBo}
	\end{equation}
which can also be obtained as a direct consequence of Archimedes' principle.

\section{Implications of the force analysis}
\label{Section:Demonstration_of_force_analysis}

\subsection{Hydrophobic cylinder}
\label{Subsection:Hydrophobic_Cylinder}

To explore some implications of the general force analysis, we consider values of $\textit{Bo}$ between $10^{-3}$ and $10^{2}$ and values of $\varGamma$ between $3/7$ and $1$. The particular value $\varGamma = 3/7$ of $\varGamma$ is the ratio of the surface tension of surfactant-laden water--air interface ($\gamma_b \approx 30$~mN/m) to that of surfactant-free water--air interface ($\gamma_a \approx 70$~mN/m), both being measured at room temperature. We use $\alpha = 1$ and $\varGamma_0 = 4/7$ in the dimensionless counterpart \eqref{surf_cont2} of \eqref{surf_cont1}, so that $\theta_b$ must lie between $70.5\degree$ and $115.4\degree$. Additionally, we fix $\theta_a$ at $115.4\degree$ and we use parentheses to indicate the particular values of $D^*$ and $Q$ corresponding to different choices of $\Gamma$. With $D$ known, we obtain the load $F_L$ required to maintain the centre of the cylinder at dimensionless vertical position $H_c$ by solving \eqref{FvFull}, \eqref{H}, and \eqref{Hc} numerically. However, in most cases we take $F_L = - F^v = 0$ and compute $H_c$. Using these results, we calculate the horizontal force components. Since we perform a static analysis, the value of the net force on the cylinder always vanishes.

\subsubsection{Implications of the vertical force analysis}
\label{Subsection:Demonstration_of_vertical_force_analysis}

\begin{figure}
\centering
\begin{overpic}{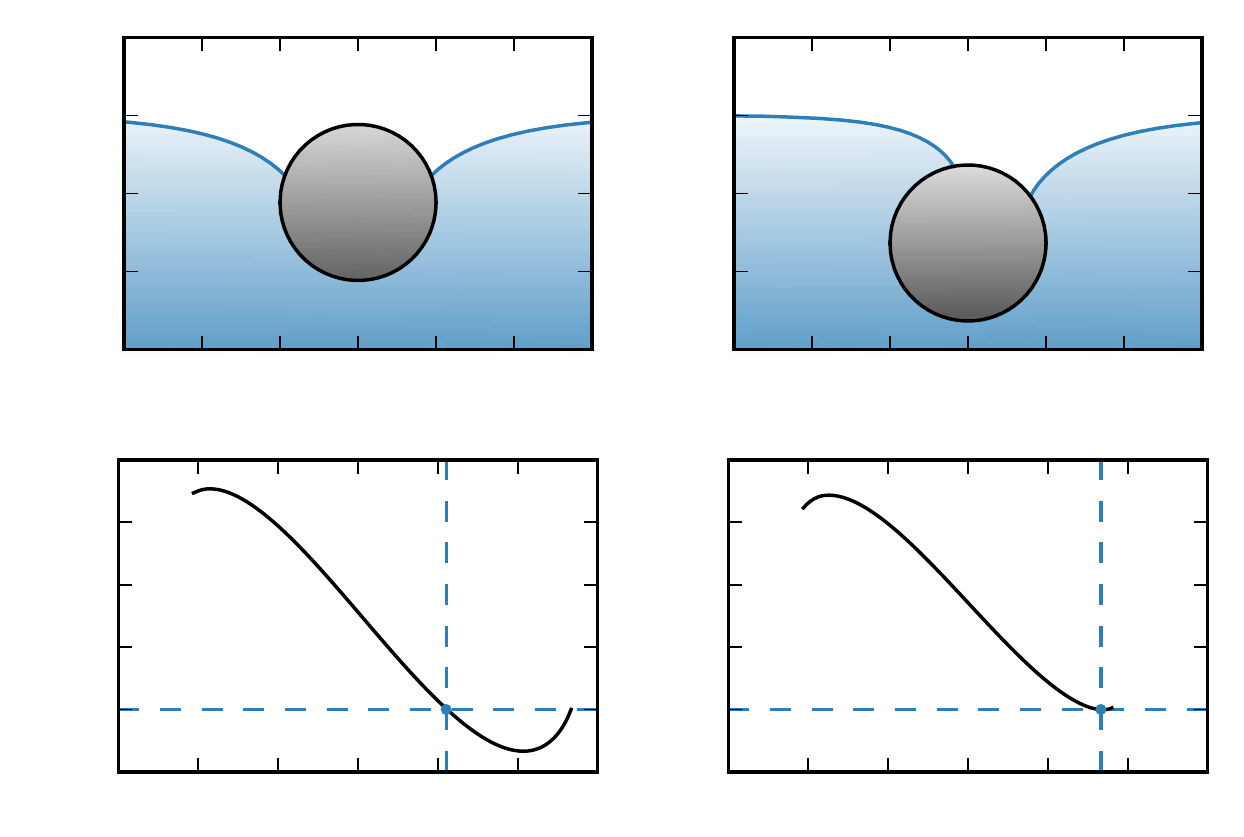}
\put(0,30.25){($c$)}
\put(50,30.25){($d$)}
%($a$)
\put(0,64.25){($a$)}
\put(24.3,51){cylinder}
\put(13,60.5){air}
\put(13,45){water}
\put(20.75,52.5){$b$}
\put(35.5,52.55){$a$}
\put(37,43){$\varGamma = 1$}
\put(3.75,51){\rotatebox{90}{$Z$}}%
\put(7.5,38.9){3}
\put(7.5,45){2}
\put(7.5,51.4){1}
\put(7.5,57.5){0}
\put(6.75,64){-1}
\put(27.5,34){$X$}%
\put(9,36.75){-3}
\put(15.5,36.75){-2}
\put(21.75,36.75){-1}
\put(28.25,36.75){0}
\put(34.5,36.75){1}
\put(40.75,36.75){2}
\put(47,36.75){3}
%($b$)
\put(50,64.25){($b$)}
\put(73.7,54){$b$}
\put(83.75,51.5){$a$}
\put(73.3,47.75){cylinder}
\put(62.25,60.5){air}
\put(62.25,45){water}
\put(85,43){$\varGamma = 3/7$}
\put(53.25,51){\rotatebox{90}{$Z$}}%
\put(56.5,38.9){3}
\put(56.75,45){2}
\put(56.75,51.4){1}
\put(56.75,57.5){0}
\put(56,64){-1}
\put(76.6,34){$X$}%
\put(58,36.75){-3}
\put(64.5,36.75){-2}
\put(70.75,36.75){-1}
\put(77.25,36.75){0}
\put(83.5,36.75){1}
\put(89.75,36.75){2}
\put(96,36.75){3}
%($c$)
\put(28,25){$\varGamma = 1$}
\put(3.5,12.25){\rotatebox{90}{$F^v = - F_L$}}
\put(6.25,5){-2}
\put(7.0,9.75){0}
\put(7.0,14.75){2}
\put(7.0,19.75){4}
\put(7.0,25){6}
\put(7.0,29.75){8}
\put(27.5,0){$H_c$}%
\put(8.5,2.75){-3}
\put(15,2.75){-2}
\put(21.25,2.75){-1}
\put(28.25,2.75){0}
\put(34.75,2.75){1}
\put(41,2.75){2}
\put(47.5,2.75){3}
%($d$)
\put(78,25){$\varGamma = 3/7$}
\put(52.5,12.25){\rotatebox{90}{$F^v = - F_L$}}
\put(55.325,5){-2}
\put(56,9.75){0}
\put(56,14.75){2}
\put(56,19.75){4}
\put(56,25){6}
\put(56,29.75){8}
\put(76.5,0){$H_c$}%
\put(57.5,2.75){-3}
\put(64,2.75){-2}
\put(70.25,2.75){-1}
\put(77.25,2.75){0}
\put(83.75,2.75){1}
\put(90,2.75){2}
\put(96.5,2.75){3}
\end{overpic}
\caption{Cross sections of two identical circular cylinders ($D = 1.63, R = 1$) at a water--air interface for respective values ($a$) $\varGamma = 1$ and ($b$) $\varGamma = 3/7$. The loads on both cylinders vanish ($F_L = -F^v = 0$). For $\alpha = 1$ and $\varGamma_0 = 4/7$, the contact angle $\theta_b$, which is located at contact line $b$, is $70.5\degree$ for $\varGamma = 3/7$ and $115.4\degree$ for $\varGamma = 1$. The contact angle $\theta_a$, which is located at contact line $a$, is fixed and set equal to $115.4\degree$. In ($a$), the cylinder is positioned higher than in ($b$) since $F_T^v$ is larger for $\varGamma = 1$ than for $\varGamma = 3/7$. Although the water--air interface profiles depicted in ($a$) are mirror images, the symmetry is broken in ($b$). In ($c$) and ($d$), $F^v$ is plotted as a function of $H_c$ for the respective cylinders depicted in ($a$) and ($b$). Movie~1 provides the corresponding cross-sections of ($c$) and ($d$). The point where the two dashed lines in ($c$) intersect the curve corresponds to the situation depicted in ($a$). At that point, $\D F^v/\D H_c < 0$, which shows that the cylinder is in a stable configuration. The point where the two dashed lines intersect the curve in ($d$) corresponds to the situation that is depicted in ($b$). At that point, $\D F^v/\D H_c = 0$, which implies that the cylinder in ($a$) has the maximum allowed mass density $D^*$ for floating.}
\label{TwoProfiles}
\end{figure}
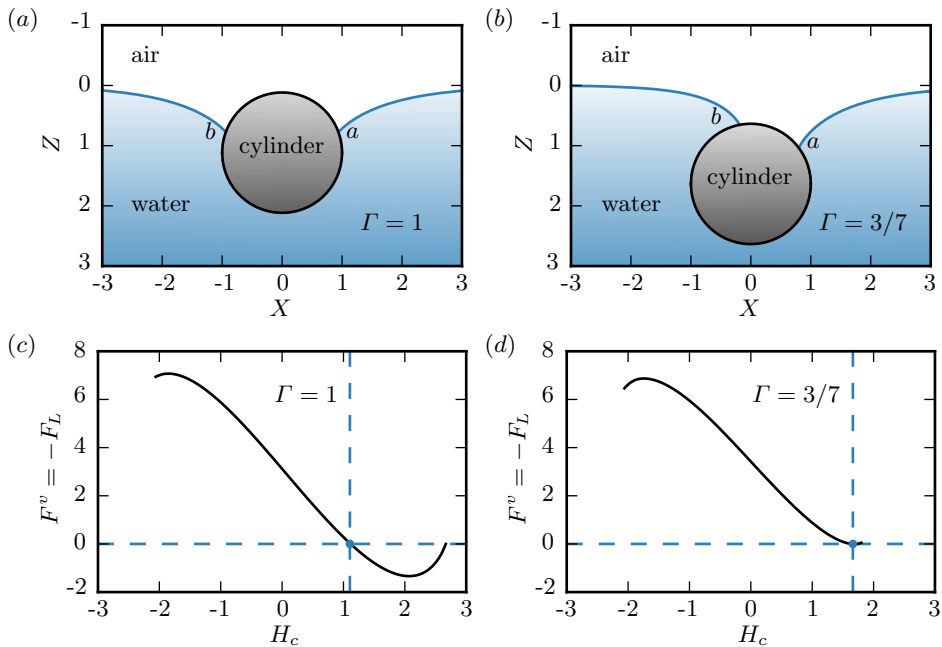

Figures~\ref{TwoProfiles}($a$) and ($b$) show cross-sections of two identical cylinders ($D = 1.63, R = 1$) at a water--air interface for respective values $\varGamma = 1$ and $\varGamma = 3/7$. The loads on both cylinders vanish, so that $F_L = -F^v = 0$. Since $F_T^v$ is larger for $\varGamma = 1$ than for $\varGamma = 3/7$, the vertical position of the cylinder in figure~\ref{TwoProfiles}($a$) is higher than that in figure~\ref{TwoProfiles}($b$). Although the water--air interface profiles depicted in figure~\ref{TwoProfiles}($a$) are mirror images, the symmetry is broken in figure~\ref{TwoProfiles}($b$). Figures~\ref{TwoProfiles}($c$) and ($d$) show $F^v$ as a function of $H_c$ for the cylinders depicted in figures~\ref{TwoProfiles}($a$) and ($b$), respectively. Movie~1 provides the corresponding cross-sections. The point where the two dashed curves in figure~\ref{TwoProfiles}($c$) intersect the solid curve corresponds to the situation that is depicted in figure~\ref{TwoProfiles}($a$). At that point, $\D F^v/\D H_c < 0$, which shows that the cylinder is in a stable configuration. However, when $\D F^v/\D H_c > 0$, the cylinder is in an unstable configuration. The point where the two dashed curves intersect the solid curve in figure~\ref{TwoProfiles}($d$) corresponds to the situation that is depicted in figure~\ref{TwoProfiles}($b$). At that point, $\D F^v/\D H_c = 0$, which implies that the cylinder in figure~\ref{TwoProfiles}($a$) has the maximum allowed mass density $D^*$ for floating.

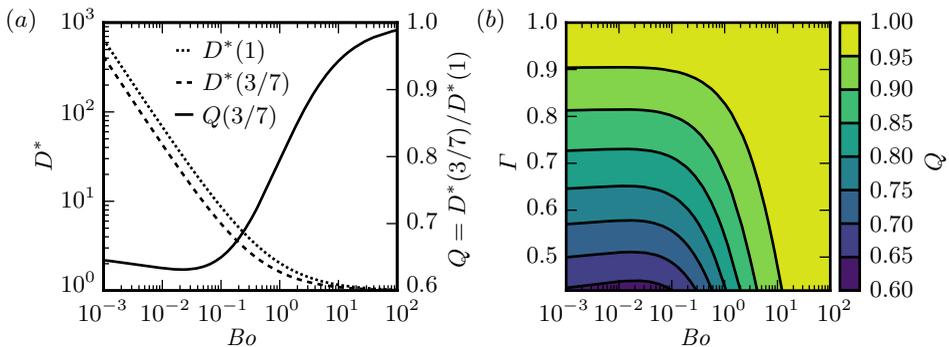
\begin{figure}
\centering
\begin{overpic}{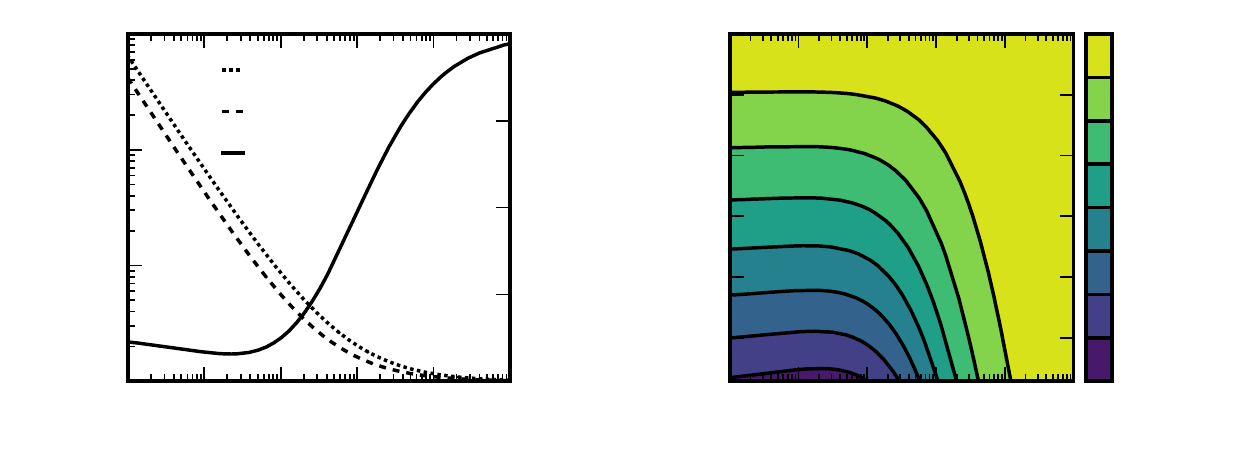}
%(a)
\put(0,33){($a$)}
\put(20.5,29.75){$D^*(1)$}
\put(20.5,26.4){$D^*(3/7)$}
\put(20.5,23){$Q(3/7)$}
\put(3,18){\rotatebox{90}{$D^*$}}
\put(46,8){\rotatebox{90}{$Q = D^*(3/7)/D^*(1)$}}
\put(23,0){$\textit{Bo}$}
\put(5.75,5.5){$10^0$}%
\put(5.75,14.25){$10^1$}
\put(5.75,23.25){$10^2$}
\put(5.75,32.75){$10^3$}
\put(7.5,2.5){$10^{-3}$}%
\put(14,2.5){$10^{-2}$}
\put(20,2.5){$10^{-1}$}
\put(27,2.5){$10^{0}$}
\put(33.25,2.5){$10^{1}$}
\put(39.5,2.5){$10^{2}$}
\put(41.75,5.5){0.6}%
\put(41.75,12){0.7}
\put(41.75,18.75){0.8}
\put(41.75,25.75){0.9}
\put(41.75,32.75){1.0}
%(b)
\put(49,33){($b$)}
\put(51.5,18){\rotatebox{90}{$\varGamma$}}
\put(95.75,18){\rotatebox{90}{$Q$}}
\put(70.25,0){$\textit{Bo}$}
\put(54.25,8.5){0.5}%
\put(54.25,13.5){0.6}
\put(54.25,18.25){0.7}
\put(54.25,23){0.8}
\put(54.25,27.75){0.9}
\put(54.25,32.75){1.0}
\put(55.5,2.5){$10^{-3}$}%
\put(61.25,2.5){$10^{-2}$}
\put(67,2.5){$10^{-1}$}
\put(73.5,2.5){$10^0$}
\put(79,2.5){$10^1$}
\put(84.5,2.5){$10^2$}%
\put(90,5){0.60}
\put(90,8.5){0.65}
\put(90,11.75){0.70}
\put(90,15.25){0.75}
\put(90,18.75){0.80}
\put(90,22.25){0.85}
\put(90,25.75){0.90}
\put(90,29){0.95}
\put(90,32.75){1.00}
\end{overpic}
\caption{($a$) $D^*(1)$, $D^*(3/7)$, and $Q(3/7)$ as a function of Bond number $\textit{Bo}$. ($b$) Contour plot of $Q$ as a function of $\textit{Bo}$ and $\varGamma$. In ($a$) and ($b$), the load is zero ($F_L = -F^v = 0$), $\alpha = 1$, and $\varGamma_0 = 4/7$. Moreover, $Q > 0.6$ for all combinations of $\varGamma$ and $\textit{Bo}$.}
\label{DmaxNonPolarSurfGraph}
\end{figure}

Figure~\ref{DmaxNonPolarSurfGraph}($a$) shows how $D^*(1)$, $D^*(3/7)$, and $Q(3/7)$ vary with $\textit{Bo}$. We observe that $D^*(1)$ is always larger than $D^*(3/7)$ and that $Q$ reaches a minimum at $\textit{Bo} \approx 2.5 \cdot 10^{-2}$. By fixing $\theta_b$ to $115.4\degree$, we find that \eqref{surf_cont2} creates this minimum. The slope of $Q$ is positive from $\textit{Bo} \approx 2.5 \cdot 10^{-2}$ since surface tension effects vanish for large values of $\textit{Bo}$. Figure~\ref{DmaxNonPolarSurfGraph}($b$) provides a contour plot of how $Q$ varies with $\varGamma$ and $\textit{Bo}$. In view of the values of $Q$ in figure~\ref{DmaxNonPolarSurfGraph}($a$), the maximum value of $\varGamma$ for each contour line is expected at $\textit{Bo} \approx 2.5 \cdot 10^{-2}$. Figure~\ref{DmaxNonPolarSurfGraph}($b$) shows that $Q > 0.6$ for all combinations of $\varGamma$ and $\textit{Bo}$.
\subsubsection{Implications of the horizontal force analysis}
\label{Subsection:Demonstration_of_horizontal_force_analysis}
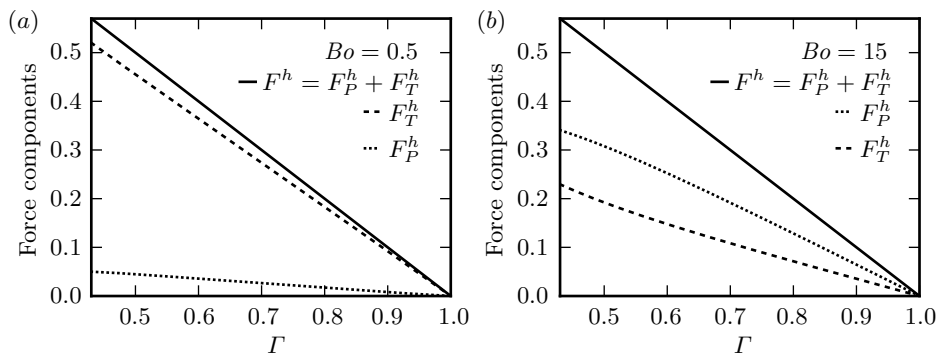
\begin{figure}
\centering
\begin{overpic}{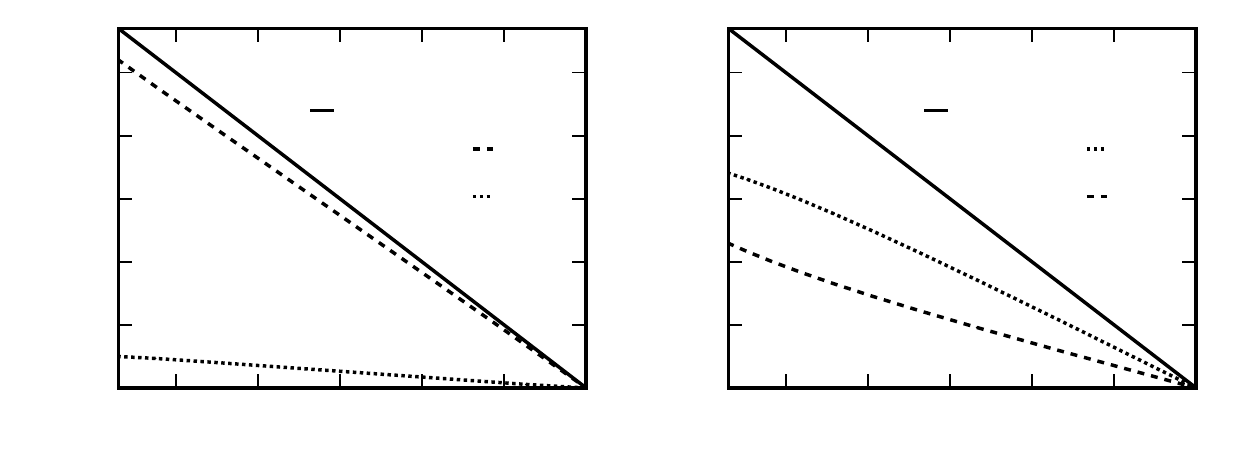}
\put(0.75,34){($a$)}%
\put(28,0){$\varGamma$}
\put(2,10){\rotatebox{90}{Force components}}
\put(34,30.75){$\textit{Bo} = 0.5$}
\put(27.5,27.5){$F^h = F_P^h + F_T^h$}
\put(40.75,24.25){$F_T^h$}
\put(40.75,20.5){$F_P^h$}
\put(12.5, 3){0.5}
\put(19.25,3){0.6}
\put(26,3){0.7}
\put(32.75,3){0.8}
\put(39,3){0.9}
\put(46,3){1.0}%
\put(5.25,5.25){0.0}
\put(5.25,10.25){0.1}
\put(5.25,15.35){0.2}
\put(5.25,20.5){0.3}
\put(5.25,25.5){0.4}
\put(5.25,30.75){0.5}

\put(50,34){($b$)}%
\put(77,0){$\varGamma$}
\put(51,10){\rotatebox{90}{Force components}}
\put(84,30.75){$\textit{Bo} = 15$}
\put(77,27.5){$F^h = F_P^h + F_T^h$}
\put(90.25,24.25){$F_P^h$}
\put(90.25,20.5){$F_T^h$}
\put(61.75, 3){0.5}
\put(68.25,3){0.6}
\put(75,3){0.7}
\put(81.5,3){0.8}
\put(88.25,3){0.9}
\put(95,3){1.0}
\put(54.25,5.25){0.0}%
\put(54.25,10.25){0.1}
\put(54.25,15.35){0.2}
\put(54.25,20.5){0.3}
\put(54.25,25.5){0.4}
\put(54.25,30.75){0.5}
\end{overpic}
\caption{The horizontal force components $F_T^h$, $F_P^h$, and $F^h$ as functions of $\varGamma$ for $F^v = 0$, $D = 0.9$, $\alpha = 1$, and $\varGamma_0 = 4/7$. In ($a$), $\textit{Bo} = 0.5$, and in ($b$), $\textit{Bo} = 15$. Additionally, the contribution of $F_P^h$ and $F_T^h$ to $F^h$ depends on $\textit{Bo}$ and $F^h  = 1-\varGamma$.}
\label{horizontal_force}
\end{figure}
We now investigate how $F^h, F_P^h$, and $F_T^h$ vary with $\varGamma$ for $F^v = 0$, considering two choices, $\textit{Bo}=0.5$ and $\textit{Bo}=0.15$, of $\textit{Bo}$ and, taking $D = 0.9$ to ensure floating. Figure~\ref{horizontal_force} shows that the contribution of $F_P^h$ and $F_T^h$ to $F^h$ depends on $\textit{Bo}$. We also observe that $F^h$ follows the simple relation
\begin{equation}    
     F^h = 1 - \varGamma.
     \label{horizontal_force_relation}
     \end{equation}
We now demonstrate that the pivotal relation \eqref{horizontal_force_relation} holds for a cylinder of arbitrary cross-sectional shape. First, using \eqref{HorizontalGeneralFinalND} and \eqref{TensionForcesND} results in
\begin{equation}
    F^h = F_P^h + F_T^h = \frac{1}{2}(H_a^2 - H_b^2) + \cos{\phi_a} - \varGamma \cos{\phi_b}.
     \label{proof1}
     \end{equation}
Substituting \eqref{H}$_1$ and \eqref{H}$_2$ in \eqref{proof1}, we next arrive at \eqref{horizontal_force_relation}, the dimensional counterpart of which shows that horizontal force $f^h$ induced by $\Delta \gamma$ is given by
\begin{equation}    
     f^h = \Delta \gamma.
     \label{horizontal_force_relationD}
     \end{equation}
The relations \eqref{proof1} and \eqref{horizontal_force_relationD} show that the surface tension difference $\Delta\gamma$ induces an asymmetry that in addition to the horizontal force due to surface tension creates a horizontal force due to hydrostatic pressure. The sum of these forces, when measured per unit length of the cylinder, is exactly equal to $\Delta \gamma$ and is independent of the vertically applied load. In \S\ref{Section:Energetic_argument} we use an energetic argument to show that \eqref{horizontal_force_relationD} is also valid for rod-like barriers with cross-sections of variable shape.

\subsection{Amphiphilic Janus cylinder}
\label{Subsection:Janus_cylinder}
The surface of a solid is amphiphilic if it consists of two subsurfaces, one hydrophilic and the other hydrophobic. If parts $\mathcal{A}_1$ and $\mathcal{A}_2$ of the surface of a cylinder of dimensionless radius $R$, described by respective coordinates $\psi \in [\psi_c - \upi, \psi_c]$ and $\psi \in [\psi_c,\psi_c + \upi]$, have different properties, then the cylinder is a symmetric `Janus cylinder'. Here, the angle $\psi_c$ describes the orientation of the cylinder. 

\begin{figure}
\centering
\begin{overpic}{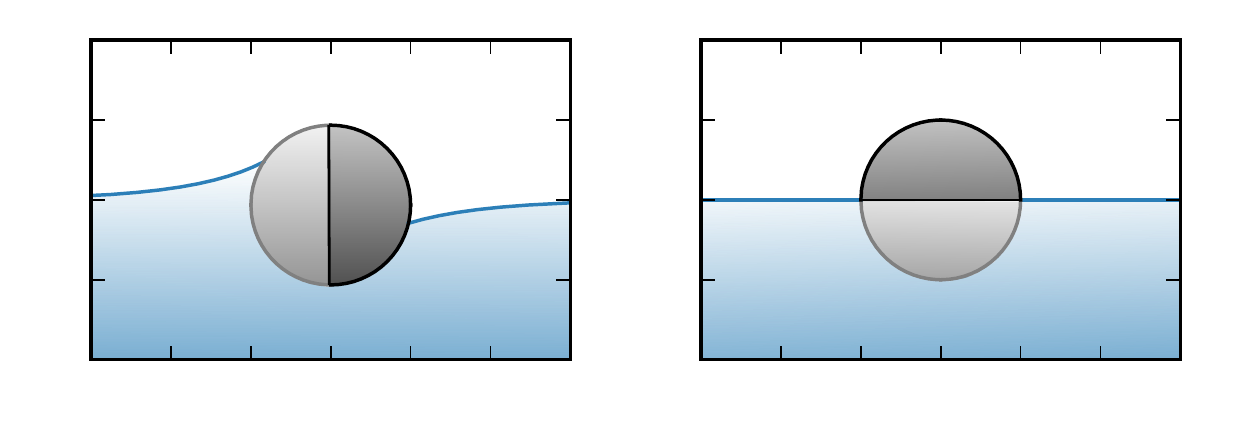}
\put(-0.5,30.25){($a$)}
\put(19.5,21.5){$b$}
\put(34,17.5){$a$}
\put(4,30){-2}
\put(4,23.5){-1}
\put(4.75,17){0}
\put(4.75,10.75){1}
\put(4.75,4.5){2}
\put(6.25,2.25){-3}
\put(12.75,2.25){-2}
\put(19.25,2.25){-1}
\put(26.1,2.25){0}
\put(32.5,2.25){1}
\put(39,2.25){2}
\put(45.25,2.25){3}
\put(25.5,-0.5){$X$}
\put(1,16.85){\rotatebox{90}{$Z$}}
\put(11,25){air}
\put(28,20.5){\rotatebox{-90}{Janus}}
\put(23,21.75){\rotatebox{-90}{cylinder}}
\put(11,9.5){water}
\put(48.75,30.25){($b$)}
\put(53.25,30){-2}
\put(53.25, 23.5){-1}
\put(54,17){0}
\put(54,10.75){1}
\put(54,4.5){2}
\put(55.1,2.25){-3}
\put(62,2.25){-2}
\put(68.5,2.25){-1}
\put(75.3,2.25){0}
\put(81.8,2.25){1}
\put(88.25,2.25){2}
\put(94.5,2.25){3}
\put(74.75,-0.5){$X$}
\put(50.25,16.85){\rotatebox{90}{$Z$}}
\put(61,25){air}
\put(72.5,19){Janus}
\put(67.5,18.8){$b$}
\put(83,18.8){$a$}
\put(71.5,15){cylinder}
\put(61,9.5){water}
\end{overpic}
\caption{Cross-sections of two identical amphiphilic Janus cylinders ($D = 0.5, R = 1$) at a water--air interface for ($a$) $\varGamma = 1$ and ($b$)  $F_L = - F^v = 0$. The respective contact angles $\theta_a$ and $\theta_b$ are located at the contact lines $a$ and $b$, and the respective orientations of the cylinder in ($a$) and ($b$) are $\psi_c = \upi/2$ and $\psi_c = 0$. The subsurfaces $\mathcal{A}_1$ and $\mathcal{A}_2$ on the dark and on the light sides of the cylinders have low ($\theta_1 = 2\upi/3$) and high ($\theta_2 = \upi/6$) degrees of wettability, respectively. Therefore, in ($a$), $\theta_a = 2\upi/3$ and $\theta_b = \upi/6$. While the dimensionless torque $T$ is equal to $1.37$ in ($a$), it vanishes in ($b$). Also, $\theta_a = \upi/2 = \theta_b$ in ($b$), so that the water--air interface is flat on both sides of the cylinder.}
\label{Janus}
\end{figure}
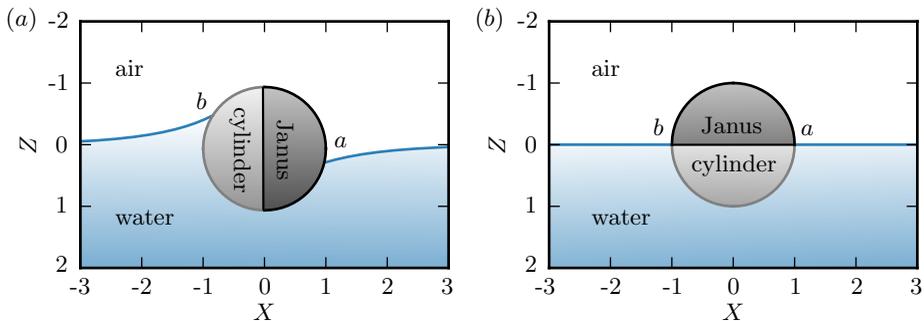

To apply our force analysis to an amphiphilic Janus cylinder, we choose $\psi_c = \upi/2$, $\textit{Bo} = 1$, $D = 0.5$, $\varGamma = 1$, and contact angles $\theta_1 = 2\upi/3$ and $\theta_2 = \upi/6$ for the respective subsurfaces $\mathcal{A}_1$ and $\mathcal{A}_2$. Figure~\ref{Janus}($a$) shows the cross-section of the Janus cylinder at a water--air interface for $F_L = -F^v = 0$. To obtain this result, we proceed as in \S\ref{Subsection:Hydrophobic_Cylinder}. Using \eqref{TorqueND} and \eqref{horizontal_force_relation}, we also obtain that a torque of dimensionless magnitude $T = 1.37$ is generated despite the absence of a horizontal force component ($F^h = 0$).

In the process of reducing the magnitude of $T_R$ to zero, the Janus cylinder rotates to ensure that a larger portion of $\mathcal{A}_2$ contacts the liquid. For $T = 0$ and $\psi_c = 0$, we use relation \eqref{FvSym} and find three regimes for which solutions can be found. These regimes correspond to the following alternatives:
\begin{enumerate}
\item $\theta = \upi/6$ and $\psi_a \geq 0$,
\item $\upi/6\le\theta\le2\upi/3$ and $\psi_a = 0$,
\item $\theta = 2\upi/3$ and $\psi_a \leq 0$.
\end{enumerate}
For $F_L = - F^v = 0$, we numerically find the solution in the second regime with $\theta = \upi/2$, meaning that the water--air interface is flat on both sides of the cylinder, as shown in figure~\ref{Janus}($b$). Using \eqref{FvSym} and \eqref{georel}, we prove that $D = 0.5$ for any amphiphilic circular Janus cylinder with $\theta = \upi/2$. Moreover, for a circular cylinder with uniform wetting properties and $\theta = \upi/2$, the liquid--gas interface is also flat if $D = 0.5$.

\cite{Casagrande1989} used an energetic argument to show that $\theta \approx \upi/2$ for amphiphilic Janus beads, with $\textit{Bo} \ll 1$, $\psi_c = 0$, and $F^v = 0$. For an amphiphilic Janus cylinder with $\textit{Bo} \ll 1$, $\psi_c = 0$, and $F^v = 0$,  \eqref{FvSym} reduces to $\phi \approx 0$. With \eqref{georel} and $\phi \approx 0$, we find that $\theta \approx \upi/2$, which shows that the result for amphiphilic Janus beads (namely $\theta \approx \upi/2$) also holds for amphiphilic Janus cylinders.

\section{Energetic argument}
\label{Section:Energetic_argument}

\begin{figure}
\centering
\begin{overpic}{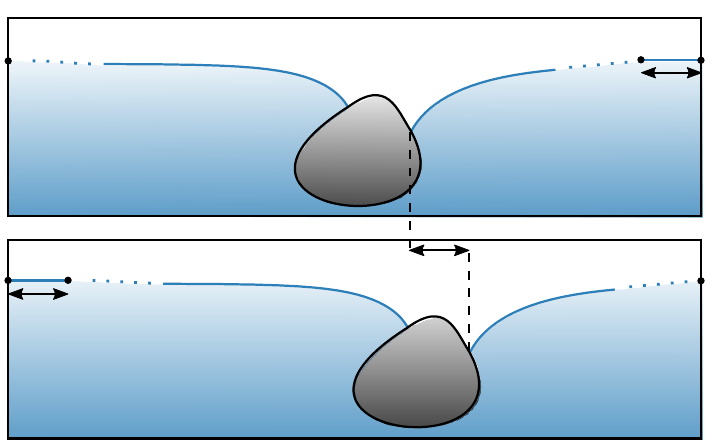}
\put(2,55.5){1}
\put(89,55.5){2}
\put(95.5,55.5){3}
\put(2,24.5){0}
\put(8.5,24.5){1}
\put(95.5,24.5){2}
\put(-7,55.5){($a$)}
\put(-7,24.5){($b$)}
\put(91.25,47.75){$\Delta x$}
\put(58.5,22.75){$\Delta x$}
\put(2.25,16.5){$\Delta x$}
\put(15,48.75){$\gamma_b$}
\put(23.5,17.5){$\gamma_b$}
\put(74,48.5){$\gamma_a$}
\put(82.5,17.5){$\gamma_a$}
\put(11.5,40){liquid}
\put(20,8.75){liquid}
\put(30,55.25){gas}
\put(38.5,24){gas}
\put(46,39){rod}
\put(54.5,7.75){rod}
\end{overpic}
\caption{($a$) Schematic cross-section of a rod-like barrier with cross-sections of variable shape, which lies at a liquid--gas interface and acts as a barrier between a surfactant-free and a surfactant-laden interface. The respective surfaces have surface tensions $\gamma_a$ and $\gamma_b$. ($b$) The rod displaced over width $\Delta x$. At points 0, 1, 2, and 3, far away from the rod, the liquid--gas interfaces are flat and located at the same height.}
\label{energy_proof}
\end{figure}
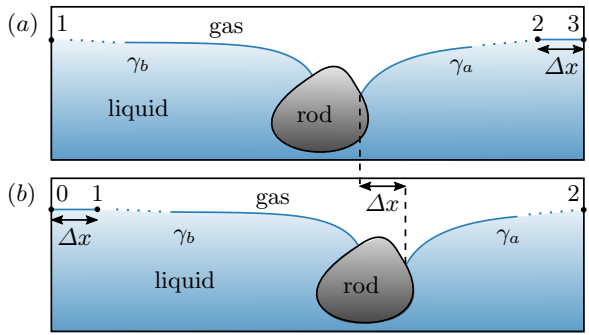

Here, we obtain \eqref{horizontal_force_relationD} with an alternative energetic argument. A single cross-section of a rod with cross-sections of variable shape is depicted in Figure~\ref{energy_proof}($a$). The rod acts as a barrier between surfactant-free and surfactant-laden interfaces with respective surface tensions $\gamma_a$ and $\gamma_b$. In Figure~\ref{energy_proof}($b$), the rod is displaced horizontally by an amount $\Delta x$ relative to its position in Figure~\ref{energy_proof}($a$).
 Let $E^{\text{($a$)}}_{13}$ and $E^{\text{($b$)}}_{02}$ denote the surface energies, both measured per unit length of the cylinder, of the system in the configurations depicted in figure~\ref{energy_proof}($a$) and figure~\ref{energy_proof}($b$), respectively. Before and after displacement, the shapes of the liquid--gas interfaces around the rod are identical and the wetting of the rod is the same. Therefore, the surface energy $E^{\text{($a$)}}_{12}$ measured between 1 and 2 in figure~\ref{energy_proof}($a$) is equal to the surface energy $E^{\text{($b$)}}_{12}$ measured between 1 and 2 in figure~\ref{energy_proof}($b$). The displacement implies that far from the rod, where the both portions of the liquid--gas interface are assumed to be flat, a surfactant-laden interface is created over width $\Delta x$ and a surfactant-free interface is consumed over width $\Delta x$. The respective energies of these liquid--gas interfaces are $E^{\text{($b$)}}_{01}$ and $E^{\text{($a$)}}_{23}$. With $E^{\text{($a$)}}_{13} = E^{\text{($a$)}}_{12} + E^{\text{($a$)}}_{23}$ and  $E^{\text{($b$)}}_{02} = E^{\text{($b$)}}_{01} + E^{\text{($b$)}}_{12}$, we express the energy difference $\Delta E$ between the interfaces depicted in figures~\ref{energy_proof}($a$) and ($b$) as 
\begin{equation}    
     \Delta E = E^{\text{($b$)}}_{02} - E^{\text{($a$)}}_{13} = E^{\text{($b$)}}_{01} + E^{\text{($b$)}}_{12} - E^{\text{($a$)}}_{12} - E^{\text{($a$)}}_{23}.
     \label{dEnergy0}
     \end{equation}
Furthermore, with $E^{\text{($b$)}}_{12} = E^{\text{($a$)}}_{12}$, $E^{\text{($b$)}}_{01} = \gamma_b \Delta x$, and $E^{\text{($a$)}}_{23} = \gamma_a \Delta x$, we see that \eqref{dEnergy0} becomes
\begin{equation}    
     \Delta E = \gamma_b \Delta x - \gamma_a \Delta x=(\gamma_b-\gamma_a)\Delta x,
     \label{dEnergy1}
     \end{equation}
from which we conclude that $\Delta \gamma$ induces a horizontal force per unit length given by
\begin{equation}    
     f^h = -\frac{\Delta E}{\Delta x} =  \Delta \gamma.
     \label{dEnergy2}
     \end{equation}
Relation \eqref{dEnergy2} proves that in general, \eqref{horizontal_force_relationD} is valid for a rod-like barrier with cross-section of variable shape.

\section{Practical implications}
\label{Section:Practical_implications}

\subsection{Rove beetle}

Certain water walking creatures that are capable of releasing surfactant from their abdominal glands, such as the rove beetle (\textit{Stenus comma}), exhibit surfactant-driven locomotion at water--air interfaces. The rove beetle supports its entire body on leg parts called \emph{tarsi}. Each of the six legs has a single \emph{tarsus}. By modeling the tarsi of such a beetle by a cylinder with a length that is equal to the combined length $l_l$ of the tarsi and by calculating the mass density of the cylinder from the weight $w_b$ of the beetle, we can use our findings to estimate the load-bearing capacity of the beetle in the absence of hydrodynamic and edge effects. From \citet{Betz2002} we know that typical values for the weight of the beetle and the total  length of its tarsi are $w_b = 26~\umu$N and $l_l = 4.7$~mm, and, moreover, that $\theta_a = 139\degree$ at $\gamma_a = 69$~mN/m, and $r = 24~\umu$m. We assume that the beetle produces a strong surfactant ($\gamma_b \approx 30$~mN/m) and that its legs remain hydrophobic ($\theta_b = 90\degree$ at $\gamma_b = 30$~mN/m). It then follows that the load-bearing capacity is 25$w_b$ for a surfactant-free water surface, 16$w_b$ for the tarsi acting as a surfactant barrier between a surfactant-laden interface and surfactant-free interface, and 10$w_b$ for a surfactant-laden interface. When the beetle walks, only three of its six tarsi might touch the water surface so that the load-bearing capacity is reduced by a factor of two. With reference to \eqref{dEnergy2}, we predict that the acceleration from rest of such a beetle is in the order of 7$g$. Since the abdominal glands of a beetle are located at the rear edge of its body, it is reasonable to assume that immediately after surfactant release the surface tension imbalance only acts on the two hind tarsi. Granted that this is so, the acceleration from rest is then reduced by a factor of three.

\subsection{Acetone droplet in a Leidenfrost state on warm water} 

To explain self-propulsion of an acetone droplet on warm water, \citet{Janssens2017} modeled an acetone droplet as a perfectly non-wetting partly submerged rigid cylinder. By ignoring hydrostatic pressure, they obtained the relation $f^h \approx \gamma_a \cos{\phi_a} - \gamma_b \cos{\phi_b}$. With the result $f^h = \Delta \gamma$ of our present findings, that relation can both be extended to account for hydrostatic pressure and simplified.

\subsection{Surface pressure measurements}

From the work of \cite{Petty1990} and the references they cite, the Langmuir balance and the Wilhelmy plate are the two devices which are most often used for measuring a surface tension imbalance. Each of these devices has its own advantages.

A Langmuir balance can be made by detaching the cylinder in figure \ref{experiment} from the walls of the water bath and connecting it to a force sensor. An accepted strategy for preventing surfactant from leaking through the gap between the walls of the bath and the edges of the cylinder involves connecting thin polytetrafluoroethylene strips to the edges of the cylinder and the bath walls \citep{Albrecht1983}. The attachment to a force sensor assures that the cylinder is essentially static and makes it possible to measure $f^hl$, where $l$ is the length of the cylinder. If the water--air interfaces are assumed to be flat, it is then evident that $f^h = \Delta \gamma$. Using this relation, the surface tension difference $\Delta \gamma$ is obtained with a Langmuir balance. The same expression, namely $f^h = \Delta \gamma$, arises from \S \ref{Subsection:Demonstration_of_horizontal_force_analysis}, while allowing both portions of the liquid--gas interface to be curved. This lends credence to the accuracy of measurements made with Langmuir balances, provided that the length and the width of the liquid bath are much greater than the capillary length determined by the liquid with the highest surface tension.

\section{Conclusions}

This work focuses on the force and torque analysis of a partly submerged circular cylinder under influence of the surface tension imbalance $\Delta \gamma$. For a hydrophobic cylinder, the vertical force analysis shows that the load bearing capacity at a water--air interface is reduced by less than 40\% after introducing a strong surfactant (resulting in a surface tension difference on the order of $\Delta \gamma = 40~$mN/m) on one side of the cylinder. From the horizontal force analysis, we learn that $\Delta \gamma$ induces an asymmetry that in addition to the horizontal force due to surface tension creates a horizontal force due to hydrostatic pressure. When measured per unit length cylinder, the sum $f^h$ of these forces is of magnitude $\Delta\gamma$. We derive this relation analytically for a cylinder of arbitrary cross-sectional shape by combining the Young--Laplace equation and a relation obtained by \cite{Keller1998}. We also consider the force analysis on an amphiphilic Janus cylinder and recover a relation obtained by \cite{Princen1969} for the vertical force on a circular cylinder due to surface tension, buoyancy, and gravity, in the special case where the surface tension imbalance vanishes, namely for $\Delta \gamma = 0$. In addition to our force and torque analysis, we use an energetic argument to show that the magnitude of $f^h$ is equal to $\Delta\gamma$ for a rod-like barrier of with cross-sections of variable shape. Finally, we discuss practical implications of the analysis for Marangoni propulsion and surface pressure measurements.

\medskip
We thank David V\'azquez-Cort\'es and Kazumi Toda-Peters for their help with the fabrication of the acrylic glass water bath. We also gratefully acknowledge support from the Okinawa Institute of Science and Technology Graduate University with subsidy funding from the Cabinet Office, Government of Japan.

\bibliographystyle{jfm}
%\bibliography{bib}

\end{document}